\documentclass[aps,twocolumn,amssymb,floatfix,preprintnumbers,amsmath,amssymb, prb]{revtex4-1}
\usepackage{epsfig}
\usepackage{graphicx}%
\usepackage{float}
\usepackage{epstopdf}
\usepackage{dcolumn}
\usepackage{bm}
\usepackage{ifthen}
\usepackage{booktabs}
\usepackage{SIunits}
\usepackage{ulem}
\usepackage{xcolor}
\usepackage{bibentry}

\begin{document}

\title{ Photo-enhanced magnetization in Fe-doped ZnO nanowires}

\author{I. Lorite}\email{lorite@physik.uni-leipzig.de}
\affiliation{Institut f\"ur Experimentelle Physik II, University
of Leipzig, Linn\'estra{\ss}e 5, D-04103 Leipzig, Germany}

\author{Y. Kumar}
\affiliation{Institut f\"ur Experimentelle Physik II, University of Leipzig,
                  Linn\'estra{\ss}e 5, D-04103 Leipzig, Germany}

                                    \author{T. Meyer}
\affiliation{Institut f\"ur Medizinische Physik und Biophysik, University of Leipzig,
                  H\"artelstra{\ss}e 16-18, D-04107 Leipzig, Germany}

                                    \author{I. Estrela-Lopis}
\affiliation{Institut f\"ur Medizinische Physik und Biophysik, University of Leipzig,
                  H\"artelstra{\ss}e 16-18, D-04107 Leipzig, Germany}

\author{P. Esquinazi}
\affiliation{Institut f\"ur Experimentelle Physik II, University of Leipzig,
                  Linn\'estra{\ss}e 5, D-04103 Leipzig, Germany}

                  \author{S. Friedl\"ander}
\affiliation{Institut f\"ur Experimentelle Physik II, University of Leipzig,
                  Linn\'estra{\ss}e 5, D-04103 Leipzig, Germany}

                  \author{A. P\"oppl}
\affiliation{Institut f\"ur Experimentelle Physik II, University of Leipzig,
                  Linn\'estra{\ss}e 5, D-04103 Leipzig, Germany}

                                    \author{T. Michalsky}
\affiliation{Institut f\"ur Experimentelle Physik II, University of Leipzig,
                  Linn\'estra{\ss}e 5, D-04103 Leipzig, Germany}

                  \author{J. Meijer}
\affiliation{Institut f\"ur Experimentelle Physik II, University of Leipzig,
                  Linn\'estra{\ss}e 5, D-04103 Leipzig, Germany}

                  \author{M. Grundmann}
\affiliation{Institut f\"ur Experimentelle Physik II, University of Leipzig,
                  Linn\'estra{\ss}e 5, D-04103 Leipzig, Germany}

\begin{abstract}

An emerging branch of electronics, the optospintronics, would be
highly boosted if the control of magnetic order by light is
implemented in  magnetic semiconductors nanostructures being
compatible with the actual technology. Here we show that the
ferromagnetic magnetization of low Fe-doped ZnO nanowires prepared
by carbothermal process is enhanced  under illumination up to
temperatures slightly below room temperature. This enhancement is
related to the existence of an oxygen vacancy V$_{\rm O}$ in the
neighbouring of an antiferromagnetic superexchange
Fe$^{3+}$-Fe$^{3+}$ pair. Under illumination the  V$_{\rm O}$ is
ionized to V$_{\rm O}^+$ giving an electron to a close  Fe$^{3+}$ ion from the
antiferromagnetic pair. This light excited electron transition
allows the transition of Fe$^{3+}$ to Fe$^{2+}$ forming stable
ferromagnetic double exchange pairs, increasing the total
magnetization. The results here presented indicate an efficient
way to influence the magnetic properties of ZnO based
nanostructures by light illumination at high temperatures.

\end{abstract}

\maketitle


Magnetic materials with  addressable electrical and/or optical properties
may add flexibility  and larger storage density to spintronics devices.
The photo-induced magnetization (PIM)
of materials is of  interest, because it would lead to the development of optospintronic devices.
The PIM phenomenon has been found below a temperature of $T \simeq 19$K
in cobalt-iron cyanide-based
Prussian blue analog\cite{Sato96}, $T \simeq 35$K for magnetic
semiconductor heterostructures\cite{Kos97} and $T \simeq 80$K for core-shell nanostructures \cite{Kos12}.
Recently, PIM after irradiation of polarized light in oxygen-deficient SrTiO$_{3}$ and at $T < 10$K
has been reported\cite{Rice14}, revealing the importance of  certain defects for a possible
manipulation of magnetization by light.
The photo induced magnetization has also been observed at room temperature, RT, in non-epitaxial film with a Nano-crystallite size of 17nm \cite{Bett09}. The results in Ref. [5] suggested that a  a stronger spin-orbit interaction given by small crystallite size samples is essential to generate the PIM in manganese zinc ferrite films \cite{Bett09}. However, it hinders the possible applications in spintronic since small spin-orbit interaction is ideally required for a larger spin coherence length.
In particular,  this is the case of ZnO based DMS materials, which are ideal candidates not only due to the large spin coherence length\cite{Matt12}, but also  because of the observed variation in the optoelectronic properties at
low magnetic fields, i.e. 0.4~T.\cite{Lori15}. In fact, studies have already been performed in  colloidal nanocrystals of Mn$^{2+}$ doped ZnO
showing a PIM property but only below
$T \sim  2~$K as shown by magnetization measurements.\cite{Stef09}
It is, therefore, of interest to further study PIM in magnetic semiconductors like ZnO
and try to get this phenomenon at higher temperatures.
In the present work, we report on the magnetic properties of low Fe-doped ZnO nanowires (NWs) under the influence of light irradiation.
Fe doping and the ionization of oxygen vacancies are found to be the key to observe PIM in nanowires
 up to $T \sim 270~$K. The PIM effect is observed only for Fe-doped ZnO deficient in oxygen.

The Fe-doped ZnO (ZF) nanowires were prepared by a carbothermal process explained
elsewhere\cite{Ale11,Lor14}. A mixture of ZnO/Fe$_3$O$_4$/Carbon in the weight proportion of 30\%/20\%/50\%
was pressed applying  4~kPa to form a pellet of 2~mm diameter. The pellet was kept in a furnace
at 1150$°$C for one hour and then cooled down to room temperature. A white foam-like powder formed by NWs was obtained on the wall  at a distance of 15~cm from the center of tubular furnace.

The magnetic moment of a bundle of NWs was measured using a superconducting
quantum interference device magnetometer (SQUID) from Quantum Design. The measurements of magnetization
under illumination were performed with a Xe-lamp coupled to the
SQUID. The light was guided by an optical fiber to the sample.
The wavelength was selected by using the corresponding filters.
The light intensity was kept low
in order not to increase the sample temperature at the
lowest measured temperatures.

Photoluminescence (PL) of the NWs at room temperature was excited
by the 325 nm line of a He-Cd KIMMON IK3301R-G laser operating
in continuous-wave mode, spectrally dispersed by a
320 mm HORIBA JOBIN YVON iHR320 monochromator and detected by a Peltier-cooled CCD camera.
X-ray diffraction measurements were performed using Philips Xpert diffractometer and
Raman spectroscopy at room temperature was performed with a micro-Raman Jobin Yvon U1000 double
spectrometer with excitation light of
532 nm and with a liquid nitrogen cooled CCD detector.
EPR measurements were carried out with a BRUKER EMX Micro X-band spectrometer at 9.41$\,$GHz.
The Q-band cw EPR spectra were recorded with a BRUKER EMX 10-40 spectrometer operating at 34$\,$GHz.
EPR spectra were simulated using the Easy Spin Matlab toolbox \cite{Stoll06}.

Scanning electron microscopy (SEM) measurements revealed that the
synthesized wires have a diameter of $\leq$130 nm and several micrometers length.
The XRD pattern of a bundle of ZF NWs along with that of undoped ZnO. The XRD results indicate the
existence of diffraction peaks of hexagonal ZnO
as a single phase with no presence of secondary phases or Fe clusters, within experimental resolution.
Particle induced x-ray emission (PIXE) measurements
were carried out to quantify the amount of Fe in the NWs. We found a Fe concentration of
$ 0.2 \pm 0.05~$at.\%. This Fe concentration is slightly smaller that
the previously reported Fe-solubility limit in ZnO \cite{Ale11}.

\begin{figure}[ht]
\centering{
\includegraphics[width=1\columnwidth]{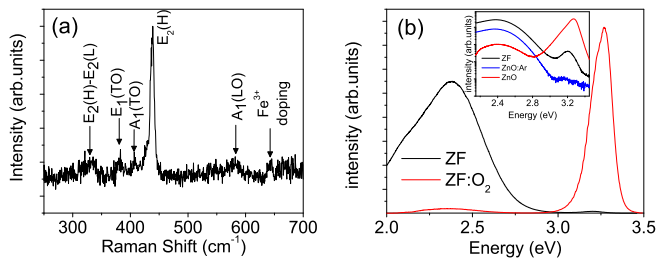}}
 \caption{a) Raman spectroscopy of a bundle of ZF nanowires. (b) Photoluminescence of ZF NWs
 as prepared (in black) and after an annealing treatment
 at 400C in oxygen atmosphere (in red).
The inset shows the photoluminescence of undoped ZnO NWs before (in red) and after Ar annealing at 400C (in blue),
 to produce oxygen vacancies, in comparison to ZF NWs (in black)}.
 \label{Raman-PL}
 \end{figure}

Raman spectroscopy allows the observation of small amount of impurities, extra phases or
dopant in ZnO\cite{Bund03,Lor09}.
In agreement with the XRD results, signatures of secondary phases have not been found  in the Raman spectrum  of the studied NWs, see Fig.~\ref{Raman-PL}, which shows the typical modes observed for ZnO\cite{Bund03}. Note that the longitudinal vibrational mode A$_{1}$(LO) at 586 cm$^{-1}$  is  related to oxygen vacancies
V$_{\rm O}$. An additional feature, observed at 644 cm$^{-1}$, has been previously related to the
incorporation of Fe$^{3+}$ ions at Zn sites \cite{Bund03}.

The existence of V$_{\rm O}$ was corroborated by PL measurements. Figure~\ref{Raman-PL}(b) shows the PL spectrum of the as grown Fe-doped ZnO NWs, which presents two different bands, i.e.
a band edge emission and a defect level emission. The band
edge emission at an energy $ \simeq$~3.2~eV (wavelength $\lambda \simeq 388~$nm)
is attributed to the recombination of free excitons and their phonon replica\cite{Shan05}.
The broad band at 2...2.75~eV (600...450~nm) represents the
defect level emission or green luminescence, which is commonly related to different type of lattice
defects  in the ZnO lattice\cite{Ale11}.
After an annealing at 400°C for the oxygenation of the NWs, the observed green luminescence
intensity is drastically reduced, see Fig. \ref{Raman-PL}(b). It suggests that  it is mostly produced by
V$_{\rm O}$'s. These vacancies are  removed by the oxygen incorporation during the
annealing treatment. It is important to mention that most of the V$_{\rm O}$'s in ZnO NWs
are defects localized at the near surface region. The  concentration of V$_{\rm O}$'s  decreases strongly
to the interior of NWs\cite{Kin13}, this facilitates its removal with low-temperature annealing treatments in
oxygen atmosphere.

\begin{figure}
\centering{
\includegraphics[width=1\columnwidth]{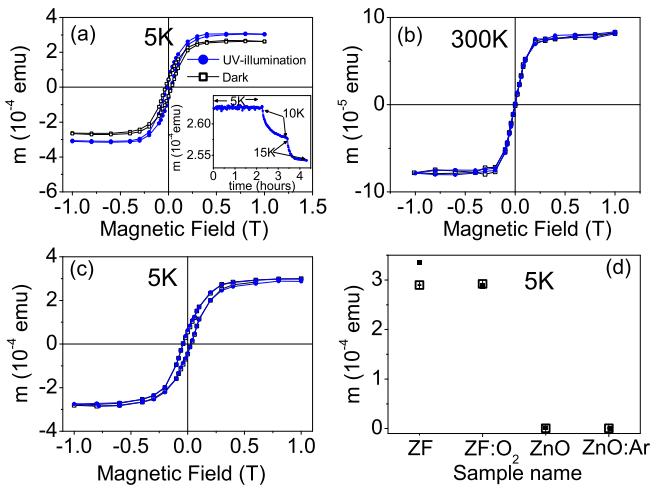}}
 \caption{ Magnetic moment vs. applied magnetic field of 6~mg of ZF NWs at (a) 5 K and (b) 300 K in dark (black squares) and under light (blue circles) at $\lambda$ = 425~nm. (c) Magnetic moment vs. applied magnetic field of 6~mg ZF NWs  in dark and under illumination (same symbols as in (a,b)) at
 $\lambda$ = 425~nm after annealing in  oxygen, ZF:O.  (d) Magnetic moment at saturation of 6mg NWs  in dark (opened squares) and under illumination (black squares) for  ZF; ZF:O2; undoped ZnO and undoped ZnO after annealing in Ar to increase the amount of O-vacancies, ZnO:Ar. The annealing treatments to obtain ZF:O2 and ZnO:Ar were performed in the previously measured ZF and ZnO respectively. The inset in (a) shows the time
independence of the PIM (after turning off the light) at 5~K and
its  exponential decrease in time after increasing the
temperature.
  The applied field was 0.1~T in this measurement.}
 \label{SQUID1}
 \end{figure}

The change of the  magnetic moment of the samples with light has been measured
in Fe-doped and undoped ZnO NWs, before and after annealing in O$_2$ atmosphere.
 Figure~\ref{SQUID1}(a,b) shows the field hysteresis of the as prepared ZF NWs, in dark and in
 light; all samples of 6~mg mass. The data here presented are
 after subtraction of a linear  in field
background. This background is due to the intrinsic diamagnetic
contribution of ZnO and the sample holder where the samples were
placed and a small paramagnetic (PM) contribution. Assuming a
Curie-law temperature dependence for the PM contribution and a
temperature independent diamagnetic one, we estimate a PM
contribution of the order of $\sim 8~\%$ of the total Fe
concentration in the ZF NWs, assuming $1~\mu_{B}$ per $Fe$ ion.

 In dark conditions all  the studied ZF NWs exhibit
 ferromagnetic ordering with a magnetic moment at  saturation of 3.1~$\times$ 10$^{-4}$~emu and 7.9 $\times$ 10$^{-4}$ emu at 5~K and 300~K, respectively (see Fig.~\ref{SQUID1}(a,b)). The measured magnetic moment at saturation would
 correspond to $\sim 0.6~\mu_{B}$/Fe (at 5~K), a value smaller than the  reported
  value in literature\cite{Kar07}. Note that the Curie temperature $T_{C} > 300$~K, in spite of the very low concentration of Fe.
This fact suggests
that regions should exist with a Fe concentration larger than the average
0.2~at$\%$  to trigger magnetic order above room temperature, likely due to double exchange.
The estimated low magnetic moment of Fe would also indicate that in addition to the FM contribution, Fe ions can also couple antiferromagneticaly (AFM) by super-exchange interaction\cite{Kar07,Kod99}, in addition to the paramagnetic contribution.
 Taking into account the value of $1~\mu_{B}$ per Fe for double exchange couple pairs\cite{Kar07,Kod99}, we estimate that $\sim 30\%$ of the Fe concentration is
 responsible for the FM signal, whereas most of the remaining Fe ions should be  AFM ordered.

After measuring the magnetic moment of the as-prepared ZF NWs in
dark, we illuminate them  with light of wavelength $\lambda$ =
425~nm. Figure~\ref{SQUID1}(a) shows the observed increase in the
magnetic moment of $\simeq 15\%$ at saturation under light and at
5~K. At this low temperature the  increase in the magnetic moment
with light has a persistent character after switching the light
off. This persistent light-driven enhancement of the ferromagnetic
moment decreases with  temperature, see  inset in
Figure~\ref{SQUID1}(a). The same set of experiments were performed
at different
 $\lambda$ = 360nm, 545nm, 690nm, however no change in the
magnetization was detected within a relative resolution of
$0.1\%$, indicating that a specific range of wavelength is needed to obtain the PIM effect

As shown previously, ZF NWs exhibit a green emission, see Fig. \ref{Raman-PL}(b), mostly due to the
existence of V$_{\rm O}$'s in the as-prepared NWs. To check for  their
role in the observed PIM, we explored the effect in the ZF NWs after annealing them in oxygen, see Fig. \ref{Raman-PL}(b), removing in this way a large part of the V$_{\rm O}$'s. After reducing the oxygen
vacancies content  we measured
the same magnetic moment as in the as-prepared NWs, in dark but  also under illumination. This result indicates that the V$_{\rm O}$'s are important only for the PIM effect. Finally, to clarify the importance of Fe doping, similar experiments were performed in non-magnetic, pure ZnO as-prepared as well as
after Ar annealing to produce V$_{\rm O}$'s. This was corroborated by photoluminescence by the observation of the  increase of the green band, in comparison to untreated pure ZnO, with an intensity comparable to the one observed in ZF NWs, see   inset in Fig. \ref{Raman-PL}(b). Ferromagnetic order in the ZnO NWs, independently of their V$_{\rm O}$'s concentrations, was observed neither  in dark conditions nor under illumination. All these results indicate that both, Fe and V$_{\rm O}$'s are necessary to observe the PIM effect in ZnO.

\begin{figure}
\centering{
\includegraphics[width=1\columnwidth]{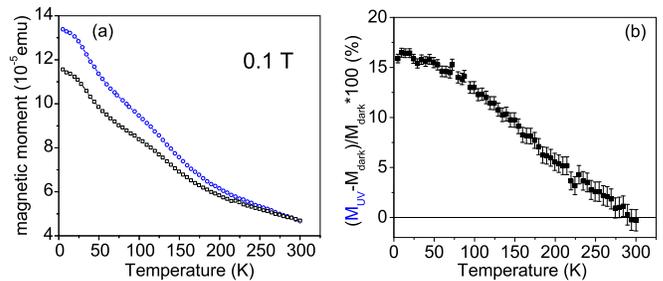}}
 \caption{ (a) Temperature dependence of ZF magnetization in dark condition (black points) and during illumination, (b) Temperature dependence of the the photo induced enhancement of the magnetization.}
 \label{SQUID2}
 \end{figure}

We measured the temperature dependence of the enhanced
magnetization under constant UV illumination. The results are
shown in Fig.~\ref{SQUID2}(a). We observe that the enhancement,
given by the difference between the results with and without
light, decreases linearly with  temperature above 50~K, see
Fig.~\ref{SQUID2}(b), vanishing at $\approx$270~K. This behavior can
be an  indication of the
existence of spin waves  in two dimensions (2D), as previously
observed in defect-induced magnetic graphite after proton
irradiation\cite{Bar07}. This apparent 2D dimensionality of the
the PIM effect in our ZF NWs agrees with the expectation that the
near surface region is the one where the PIM effect originates.

\begin{figure}
    \centering{
\includegraphics[width=1.\columnwidth]{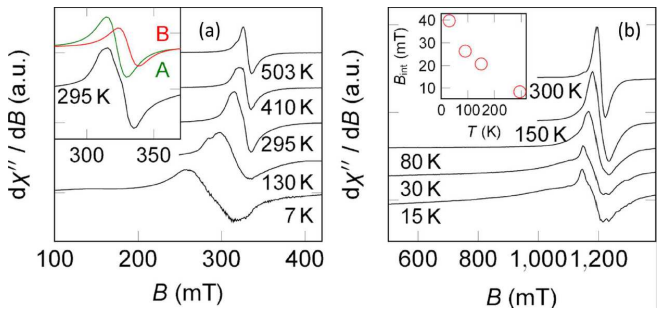}}
    \caption{(a) Experimental EPR spectra at X-band frequency of ZF NWs at temperatures from 7$\,$K to 503$\,$K. The inset shows the two components of the room temperature signal. (b) Q-band EPR spectra of ZF nanowires at temperatures between 15$\,$K and room temperature. The inset shows the calculated internal fields for species A.}
    \label{EPR}
\end{figure}

To further clarify the mechanisms behind the  magnetic order and
the PIM origin observed in the ZF NWs, electronic paramagnetic
resonance (EPR) measurements were performed between 7~K and 503~K.
In Fig.~\ref{EPR} EPR measurements of a bundle of ZF NWs  at
X- and Q-band frequencies are presented. At room temperature the X-band measurements show a
broad signal around $g=2$. Attributing these signals to Fe$^{3+}$ ions we
suggest that Fe$^{3+}$ is not magnetically diluted and magnetic dipolar and exchange interactions between Fe-ions lead to a coalescence of the fine structure pattern of Fe$^{3+}$. The signal  can be
simulated by two main species, A and B,  see inset in Fig.
\ref{EPR} (a), which are observed in a wide temperature range. The
X-band measurements show an apparent $g$-value increase from
2.03 at 503~K to about 2.35 at 7$\,$K. To study the nature of the
EPR signals, Q-band measurements were conducted, see Fig.
\ref{EPR}(b). Here we observe a slight increase of the $g$-factor from 2.035 at room
temperature to 2.06 at 15$\,$K, suggesting the presence of
temperature dependent internal fields which can be estimated from the comparison of
the X-band and Q-band spectra \cite{Fain99} and are shown  in the
inset of Fig.~\ref{EPR}(b). This allows us to recalculate the real value of
$g = 2.00(1)$ at room temperature in good agreement with the
value for Fe$^{3+}$.\cite{Fain99} Additionally, we observe that the line width increases
 towards lower temperatures, see Fig.~\ref{EPR}(a), suggesting that the Fe ions observed by EPR do not couple ferromagnetically\cite{Sch94} but an antiferromagnetic coupling exists.
With the presence of internal fields attributed to Fe-ions ferromagnetically coupled, the overall EPR results support the interpretation of
the magnetization that both FM and AFM ordered Fe-ions
co-exist in the studied NWs \cite{Win72}.

As noted above, to explain  the observed magnetic order at such high temperatures and with such
small Fe concentration, we need to assume  the formation of a non
homogeneous distribution of Fe-ions which allows the co-existence of FM, Fe$^{2+}$-Fe$^{3+}$, and AFM regions, Fe$^{3+}$-Fe$^{3+}$.
Moreover, the  V$_{\rm O}$'s can be placed in the neighboring  of
 an AFM couple.
 By light illumination  the mention V$_{\rm O}$ can be ionized to form V$_{O^+}$. When the V$_{\rm O}$ is close to an Fe$^{3+}$, this
 photo-ionization can produce an electron transfer from the V$_{\rm O}$ to the nearby Fe$^{3+}$ ion transforming it into  Fe$^{2+}$.
 This will produce Fe$^{2+}$-Fe$^{3+}$ FM couple pairs increasing the density of double exchange couples
that contribute to the FM order in detriment of the pairs AFM
ordered.

The found PIM effect in ZnO doped with a nominally low
concentration of Fe and our interpretation based on the
light-induced ionization of the   oxygen vacancies near Fe ions,
localized mainly at the surface of the NWs, shed light to the possibility
to explore light-induced magnetism  in semiconductors at much
higher temperatures than the ones reported previously. The effect
found in our study  could be applied to other type of oxides in
which the ionization of vacancies by light is possible. The PIM
effect and the possibility to have it in nanostructures as small
or smaller than in our NWs, should substantially support a further
development of magnetic semiconductors for optospintronics
application at technologically relevant wavelengths.

\begin{acknowledgments}
This work  was partially  supported  by CIUNT under Grants 26/E439
and 26/E478, by ANPCyT-PICTR 35682, and by the Collaborative
Research Center SFB~762 ``Functionality of Oxide Interfaces''. We
are grateful for the support within the DFG priority program
SPP~1601 ``New Frontiers in Sensitivity for EPR Spectroscopy''.
\end{acknowledgments}

\bibliographystyle{apsrev4-1}

%

\end{document}